\documentclass[a4paper,fleqn,usenatbib]{mnras}
\usepackage[T1]{fontenc}
\usepackage{ae,aecompl}

\usepackage{graphicx}	
\usepackage{amsmath}	
\usepackage{amssymb}	

\usepackage{gensymb}

\title{Observing the polarisation pattern of Saturn using CARMA}

\author[M. Aich et al.]{\parbox{\textwidth}{\raggedright Moumita Aich$^{1,2}$\thanks{E-mail: aich@ukzn.ac.za}, Kenda Knowles$^{1}$, Kavilan Moodley$^{1}$, Jonathan Sievers$^{2,3}$, Matthew Hedman$^{4}$}
\vspace{0.4cm}\\
\parbox{\textwidth}{\raggedright 
$^{1}$~Astrophysics \& Cosmology Research Unit, School of Mathematics, Statistics \& Computer Science, University of KwaZulu-Natal, Durban 4041, SA\\
$^{2}$~Astrophysics and Cosmology Research Unit, School of Chemistry and Physics, University of KwaZulu-Natal, Durban 4041, South Africa  \\
$^{3}$~National Institute for Theoretical Physics (NITheP), University of KwaZulu-Natal, Private Bag X54001, Durban 4000, South Africa \\
$^{4}$~Physics Department, University of Idaho, Moscow, ID 83844, USA
}
}

\begin{document}
\maketitle
\label{firstpage}

\begin{abstract}
We observe Saturn and its ring system at wavelengths of 1.3 mm (220 GHz) using the Combined Array for Research in Millimeter-wave Astronomy (CARMA) interferometric array. We study the intensity and polarisation structure of the planet and present the best polarisation data of Saturn at these frequencies. Observations using CARMA E-array configuration exhibited some anomalous polarisation pattern in the rings. We provide details of our analysis and discuss the possibility of self gravity wakes in Saturn's ring system resulting in this anomaly. We observe Venus in intensity and polarisation to cross-check the levels of polarisation signal detectable by CARMA. We also discuss how limitations in CARMA instrumental accuracy for observing weakly polarised sources, project this signature as an upper bound of polarisation measurements of Saturn using CARMA.
\end{abstract}



\section{Introduction}
Experiments aimed at making high-precision measurements of the cosmic microwave background (CMB) anisotropy require accurate knowledge of the angular response of the instrument and absolute calibration to a few percent level accuracy or better. For experiments targeting the CMB polarisation signal, the polarisation fraction and polarisation orientation must also be measured to high accuracy. Calibration of these latter measurements necessarily depends on a bright and well-measured polarised source. It is common for CMB experiments, both ground-based and space-based, to use bright sources such as planets or fixed celestial sources for absolute calibration and to measure beam profiles. Time variability is a key factor for which many of the stronger extra-galactic sources such as quasars are rejected as calibrators. After correcting for slowly varying seasonal effects, planets are good candidates as calibration sources since they provide a signal with suitable intensity and high stability. The calibration of the polarisation amplitude and orientation presents more of a challenge as non-variable and compact celestial calibrators that are sufficiently polarised are not common at millimeter wavelengths. 

The Atacama Cosmology Telescope (ACT) which is a six-meter diameter telescope in the Atacama Desert of northern Chile, is designed to make high-resolution measurements of the CMB anisotropies, observing at frequencies of 148, 218 and 270 GHz . For the case of the second-generation, polarisation-sensitive receiver ACTPol, which has arc-minute resolution, a bright and well-measured compact polarised source is necessary for polarisation calibration. Saturn is bright enough to be detected as a microwave source \citep{weiland}. Its brightness and relatively high degree of polarisation that arises due to scattering of Saturnian radiation by the rings, makes it suitable as a polarisation calibrator \citep{grossman, tak}. However, since the polarisation fraction changes along with the viewing cross-section of Saturn's rings from Earth as it moves about the sky, our primary aim was to develop a model for Saturn as a polarisation calibrator at 1 mm wavelengths. If successful, we anticipated using Saturn as the primary polarisation calibrator for ACTpol. Surprisingly, there are no good polarisation data on Saturn at mm wavelengths. The goals of this project were mainly to study the polarisation structure of Saturn for use as a calibrator, particularly for CMB experiments which in turn would require constraints on scattering properties of the disk material in Saturn's rings. 

Observations from the Combined Array for Research in Millimeter-wave Astronomy \href{http://www.mmarray.org/}{(CARMA)} would have also allowed interesting planetary science. These observations provide a direct means of probing the atmosphere of Saturn and understanding the composition and distribution of particles in Saturn's rings at a wavelength comparable to individual ring particle sizes. At long wavelengths, the rings' brightness is dominated by scattered radiation from Saturn but at shorter wavelengths, thermal emission from the rings themselves make a large contribution \citep{dunn}. Polarisation data provides a way to measure the relative strength of these two components. 

If the rings were composed exclusively of isolated, spherical particles, then the thermal emission should not have a strong net polarisation. In this case, the polarised component in the rings' millimeter-wave emission would come entirely from the scattered Saturn-shine, and determining the relative contributions of the two components would be relatively straightforward. However, the real situation will likely be more complicated since particles in Saturn's dense rings are not isolated. Indeed, in many parts of the A and B rings the particles are organized into elongated aggregates known as self-gravity wakes \citep[][and references therein]{colwell,nicholson,hedman}. These could influence the polarisation of both the thermal and the scattered radiation The key point is that the orientation of the wakes is not radial, so the polarised signal from the wakes should be different on the two sides of the ring. Multi-frequency datasets would  be useful to separate the wake signal from the particle scattering/thermal signal. 

In the following sections we discuss our observation setups, analysis and the results of the two set of observations using CARMA separated by a year. The results from the observations depict some peculiarity in the polarisation pattern on the ring system. Physical structures like self-gravity wakes could be attributed for this discrepancy. However the level of polarisation detected is on the edge of the dynamic range limit for CARMA where errors in leakage calibrations inherent in the instrument, could give false detection of polarisation. Further analysis would be needed to distinguish between the effect being physical or purely instrumental. Unfortunately because of the decommissioning of CARMA it was not possible to obtain additional data. 


\section{Observations}CARMA is a university-based interferometer consisting of six 10.4-meter (OVRO array), nine 6.1-meter (BIMA array), and eight 3.5-meter antennas that are used in combination to image the astronomical universe at millimeter wavelengths. Located at a high-altitude site in eastern California, CARMA provides a combination of sensitivity, broad frequency coverage, sub-arcsecond resolution and wide-field heterogeneous imaging capabilities, along with innovative technologies and educational opportunities. This is an aperture synthesis array, typically operating as two independent sub-arrays of 15 and 8 antennas, respectively. In the CARMA-15 sub-array, there are two receiver bands, 3 mm and 1 mm, and the spectral line correlator. A basic aperture synthesis observation makes an image the size of the the primary beam $(\lambda/D \sim 1 \arcmin$ at 100 GHz; $0.5 \arcmin$ at 230 GHz) with a resolution corresponding to the maximum separations of the antennas. Since polarisation is unavailable at 3mm, we have used CARMA 1mm E-array configuration (using the 10m and 6m antennae, this is the most compact configuration) to observe Saturn. Each of the antennae are equipped with two 1 mm receivers which are individually sensitive to left circular polarisation (LCP) and right circular polarisation (RCP). We had two sets of observations as mentioned in the subsections below. 
																															
\subsection{Observation set I}\label{obs1}
\begin{figure}
\centering
\includegraphics[width=9cm]{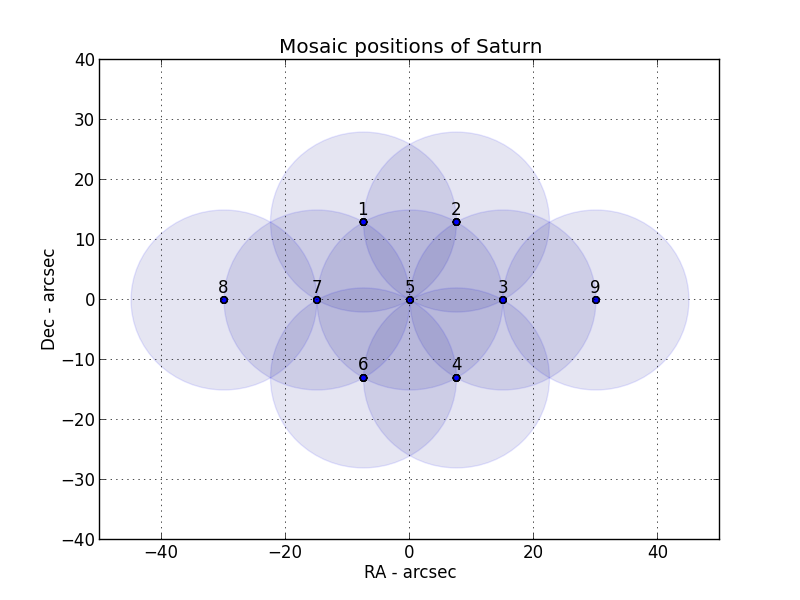}
\caption{Nine mosaic positions of Saturn: A central pointing along with a hexagonal mosaic to span Saturn's disk. The shaded circles show the size of the primary beams of the 10-m telescopes. 
Distance from the central mosaic to the outer mosaic points is $15\arcsec$. Two mosaic points along each side of the hexagon spans the rings with an angular distance of $1\arcmin$.}\label{mosaic}
\end{figure}

\begin{figure}
\centering
\includegraphics[width=9cm]{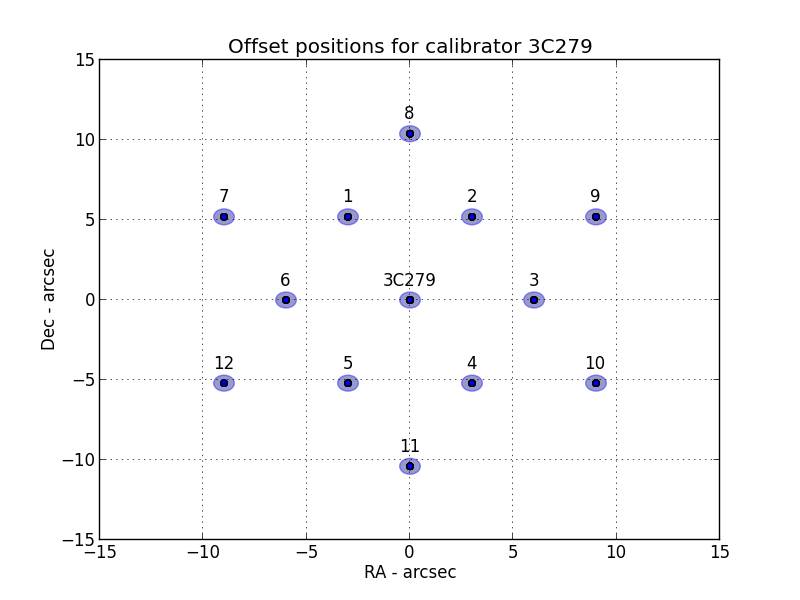}
\caption{Twelve offsets for the calibrator 3C279: Six offset pointings are arranged along an inner hexagon at an angular distance of $6\arcsec$ from the central pointing. The other six offset pointings are arranged along an outer hexagon at an angular distance of  $12\arcsec$ from the central pointing.}\label{offset_pos}
\end{figure}

 The first set of observation was done during 17 July, 2012. Saturn was at  ($\alpha$, $\delta$) = (13:28:49.91, -6:38:26.09). The quasar 3C279 at ($\alpha$, $\delta$) = (12:56:11.17,-5:47:21.52) with a flux of 12.4 Jy at 220 GHz, was close and bright enough to be used as our phase and gain calibrator. To observe the polarisation pattern across the disc and if possible on the rings, we use an extended mosaic pattern for Saturn as shown in Figure \ref{mosaic}. We choose a seven-point hexagonal mosaic pattern along the disc of Saturn aligned with the rings, one at the central position and one on either side of the disc, oriented along the axis of rings. This nine-point mosaic pattern is based on arranging the beam pattern for the largest 10 m CARMA antennae (with FWHM $\sim 30 \arcsec$) over the field of view. In order to measure the instrumental polarisation, we observe the calibrator at twelve offset positions, as shown in Figure \ref{offset_pos}. However, due to an error in the observing script, only two of the offsets were observed at a position that was useful to measure the instrumental polarisation. The total observation time was 3.75 hours for the entire track, concentrating 2.53 hours for Saturn and 0.29 hours for the calibrator 3C279. Each of the twelve calibrator offsets had a total track of $\sim 0.08$ hours. 

\subsection{Observation set II}\label{obs2}
The second set of observation was done during 31 July, 2013. Saturn was at  ($\alpha$, $\delta$) = (14:14:25.41, -10:58:50.48). The same quasar 3C279 was close and bright enough to be used as our phase and gain calibrator. However, we reduced the complexity of the mosaic pattern for the source and used a three point mosaic for Saturn, one at the center and two on either side of the ring. Ideally we need to have offset pointing for the calibrator too which would help in the final data analysis to correct for the instrumental polarisation. Due to complications involved and limited observing time, we avoided this during this time slot. The total observation time was 3.54 hours, concentration 2.79 hours for Saturn and 0.75 hours for the calibrator 3C279. 

\subsection{Auxiliary observations}\label{dynamic}
Limitations on the accuracy of polarisation measurements made with CARMA include signal to noise effects, uncertainties in polarisation leakage corrections and primary beam polarisation. For a bright source which is weakly polarised, errors in the leakage calibrations can give false detection of polarisation, thus providing a \textit{``dynamic range limit''} for CARMA polarisation measurements. Also for extended sources which are not perfectly on-axis aligned, the polarisation varies across the primary beam of the telescope. To test this we performed a third set of observations with CARMA on 4 August, 2013 using another cloud-shrouded planet - Venus. This will act as a supplementary set of data to check for instrumental polarisation leakages. The observation script uses 3C279 as the `source' and Venus at ($\alpha$, $\delta$) = (11:09:50.25, 6:41:35.35) and Saturn at ($\alpha$, $\delta$) = (14:15:02.97, -11:03:11.80) as the `calibrators'. The script was executed long enough to get a good leakage solution from 3C279. Venus during the observation was $\sim 12\arcsec$ in diameter, similar to the diameter of Saturn's disk. Venus was observed for the first part of the track and Saturn for the second part. Due to time constraint, we made single field images of the targets and used only a central mosaic on Saturn without any mosaics along the rings. This set of observation was performed as a cross-check for polarisation measurement using the CARMA E-array. 

\section{Data Reduction}
 We used Miriad for the reducing the data from CARMA. Apart from standard flagging, applying CARMA baseline solutions and usual passband and gain calibrations, polarisation measurements requires two additional calibrations -- XY phase and leakage. The XY phase is the LCP-RCP channel phase difference, caused by delay differences in the receiver and systematics in cabling. XY phase calibration errors rotate the position angles of linear polarisation vectors on the sky. For circularly polarised feeds, changing the XY phase by 180\degree rotates polarisation vectors by 90\degree. 

We derive calibration solutions for the polarisation data using the Miriad program \textit{xyauto}. The XY phase calibration is done using polarised noise sources in the cabins of the 10-m telescopes (rather than observations of polarised astronomical sources). After the usual passband calibration, all RCP phases are zero because RCP is the reference polarisation.  LCP phases are nonzero only for the 10 m antennas because only these antennae are outfitted with polarised noise sources. The aim is to apply a passband correction that is needed to make the LCP phases equal the RCP phases on a linearly polarised noise source.  Initially developed for the \href{http://tadpol.astro.illinois.edu/}{Telescope Array Doing polarisation (TADPOL)} project of CARMA, \textit{xyauto} derives the XY phase corrections by fitting a passband to the LR auto-correlation spectra and the details are outlined in \cite{hull}. On rewriting the data with \textit{uvcat} to apply the XY phase correction, the LCP and RCP channels will be aligned in phase for all 10m spectra. In order to align the LCP and RCP phases for the 6m antennas, it is essential that one of antennas 10m antennae be used as the reference antenna when solving for the passband fit using \textit{mfcal}. Since the L-R phase difference is zero for every channel of the reference antenna spectrum, and since the normal passband correction ensures that all LL and RR cross-correlations have zero phase with respect to the reference antenna, applying the regular passband transfers the XY phase correction to all other antennas. This ensures the LCP vs RCP phase correction is not lost, the polarisation leakages will be accurate and the polarisation position angles will not be rotated by arbitrary angles.

Cross-coupling between the L and R channels from each antenna due to systematics in polarisers makes polarisation leakage correction an essential step in polarisation observations. Since the CARMA receivers are fixed, the instrumental polarisation leakages are stable in time and reproduce closely daily.  LR and RL cross-correlated signals from a polarised source vary as the parallactic angle changes and hence it is easy to distinguish it from leakage effects. We use the MIRIAD program \textit{gpcal} to derive the leakage corrections from observations of a bright source (we use the calibrator 3C279) that is observed over a wide range of parallactic angle. This program is effective irrespective of the polarisation characteristics and/or polarisation state of the leakage calibrator.

 After the passband correction and leakage calibration, we generate maps from the fully calibrated visibilities using the MIRIAD task \textit{invert}. Since we use nine source mosaics for Saturn, we use the \textit{mosaic} option to process multiple pointings and generate a linear mosaic of these pointings. For the deconvolution process, instead of using the standard \textit{clean} algorithm, we have used \textit{mossdi} deconvolution algorithm which is usually utilized for mosaiced image of extended sources.  We perform iterative phase-only \textit{selfcal} on Saturn. However, instead of using an iterative model for Saturn during \textit{selfcal}, we use the \textit{apriori} option. This task in mainly used if the source in the visibility data is either a planet, or a standard calibrator; this causes the model data to be scaled by the known flux of the source. In this case \textit{selfcal} does not read in a separate file containing a model, but assumes that the planet is a disk with uniform brightness temperature.  The major and minor disk diameters and the planet brightness temperature are written into the Miriad header by the online system. It is worth noting that the Saturn model does not include the rings. For a planet, this flux will be a function of baseline; in addition this option should also be used for a phase selfcal, to get the correct weighting of the different baselines in the solution. Using these new calibration solutions, we generate the final Stokes I, Q and U maps of the source.

\section{Instrumental polarisation and systematics}
In the following subsections, we discuss briefly about instrumental polarisation leakages and other systematics. 


\subsection{Correcting leakage using the calibrator offsets}\label{leakage-corr}
\begin{figure*}
\centering
\includegraphics[width=4.5cm,angle=-90]{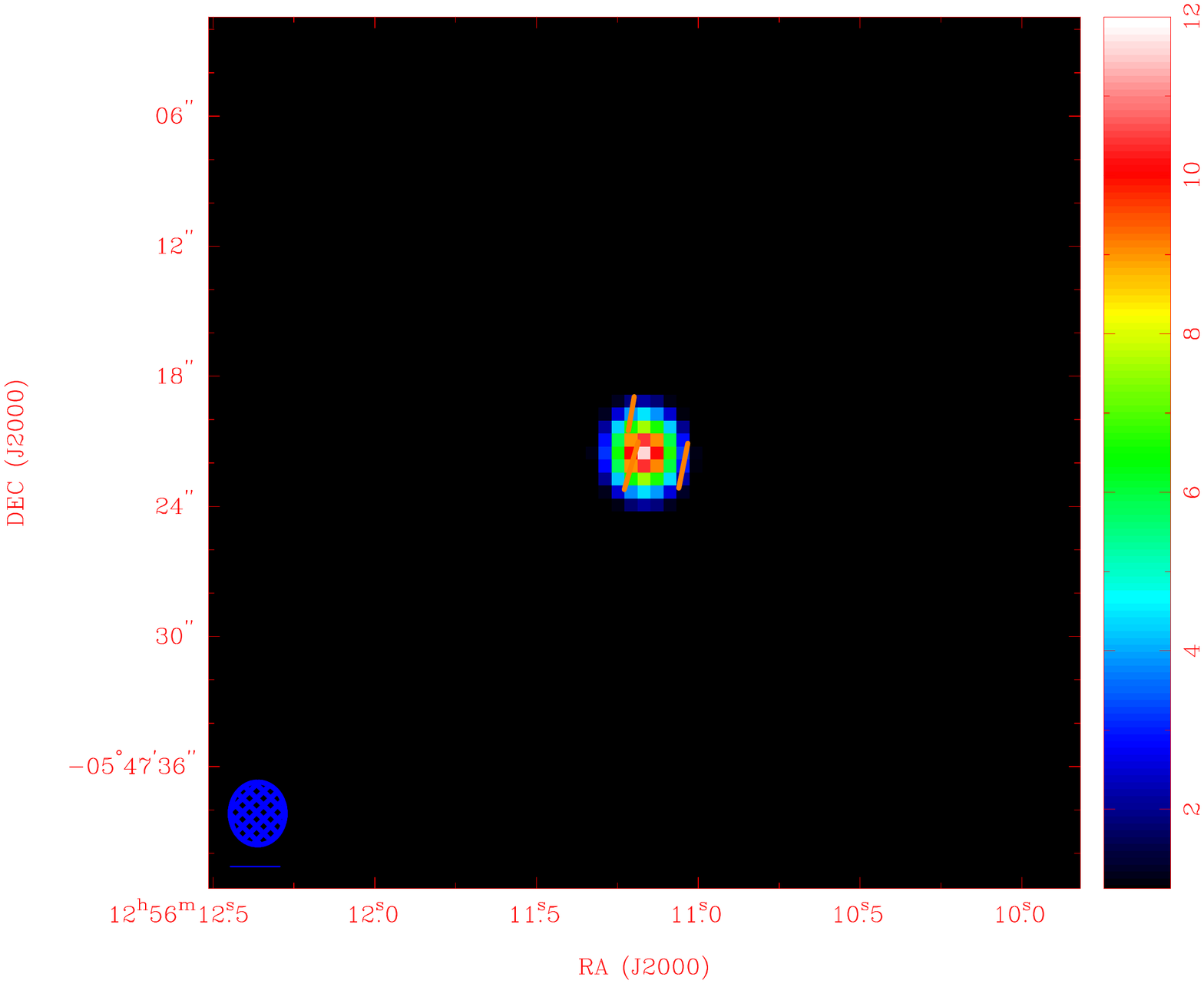}
\includegraphics[width=4.5cm,angle=-90]{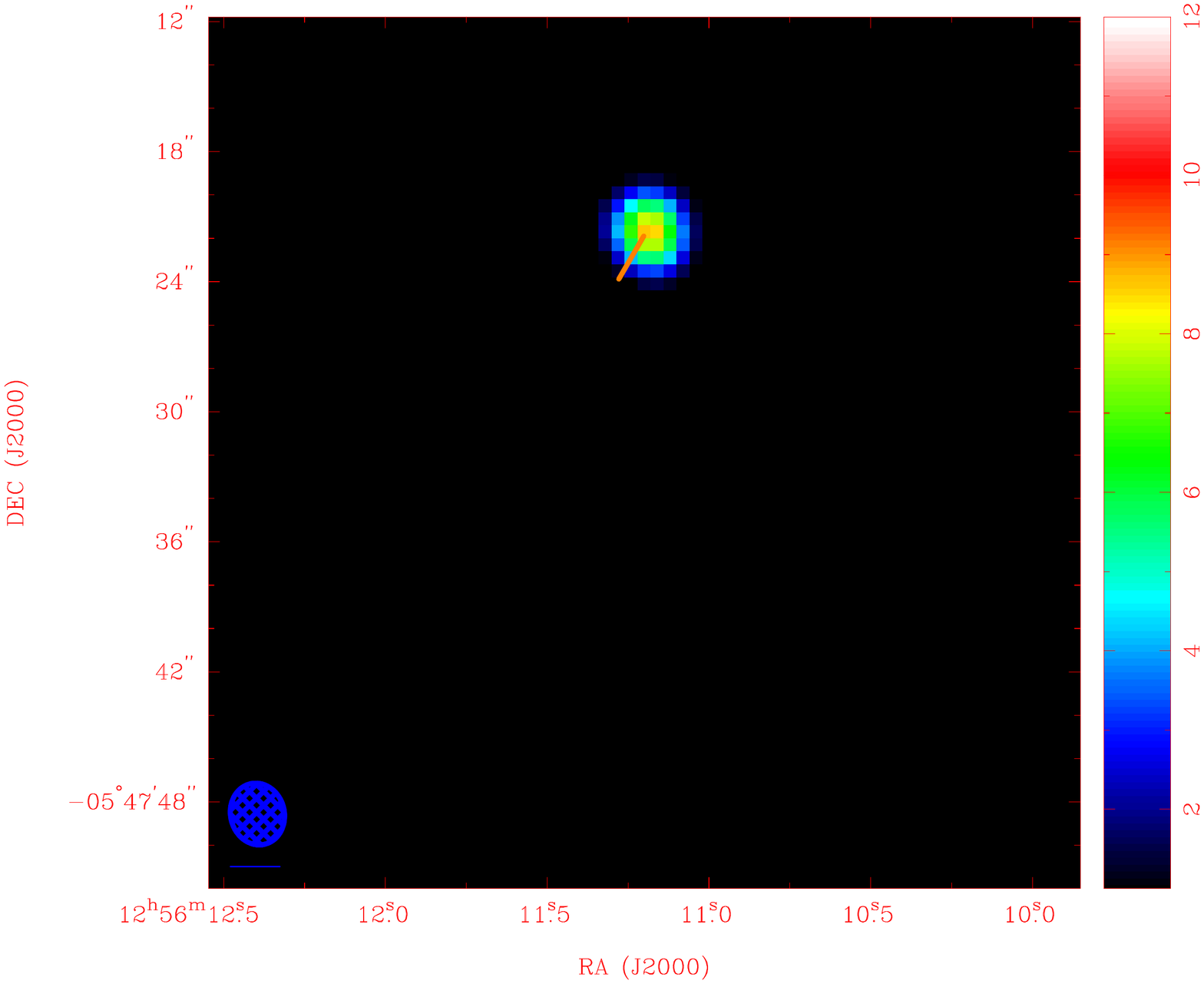}
\includegraphics[width=4.5cm,angle=-90]{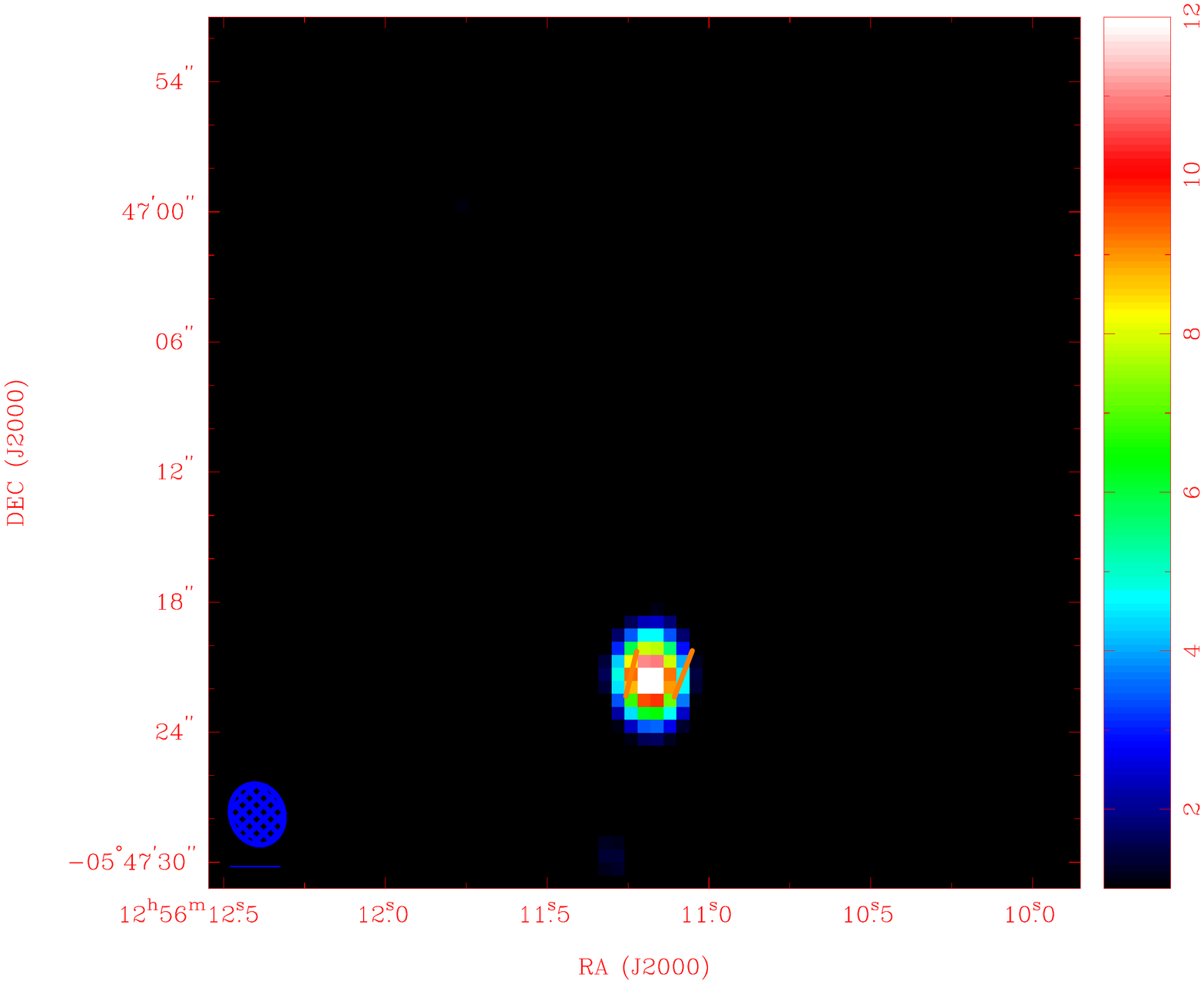}
\includegraphics[width=4.5cm,angle=-90]{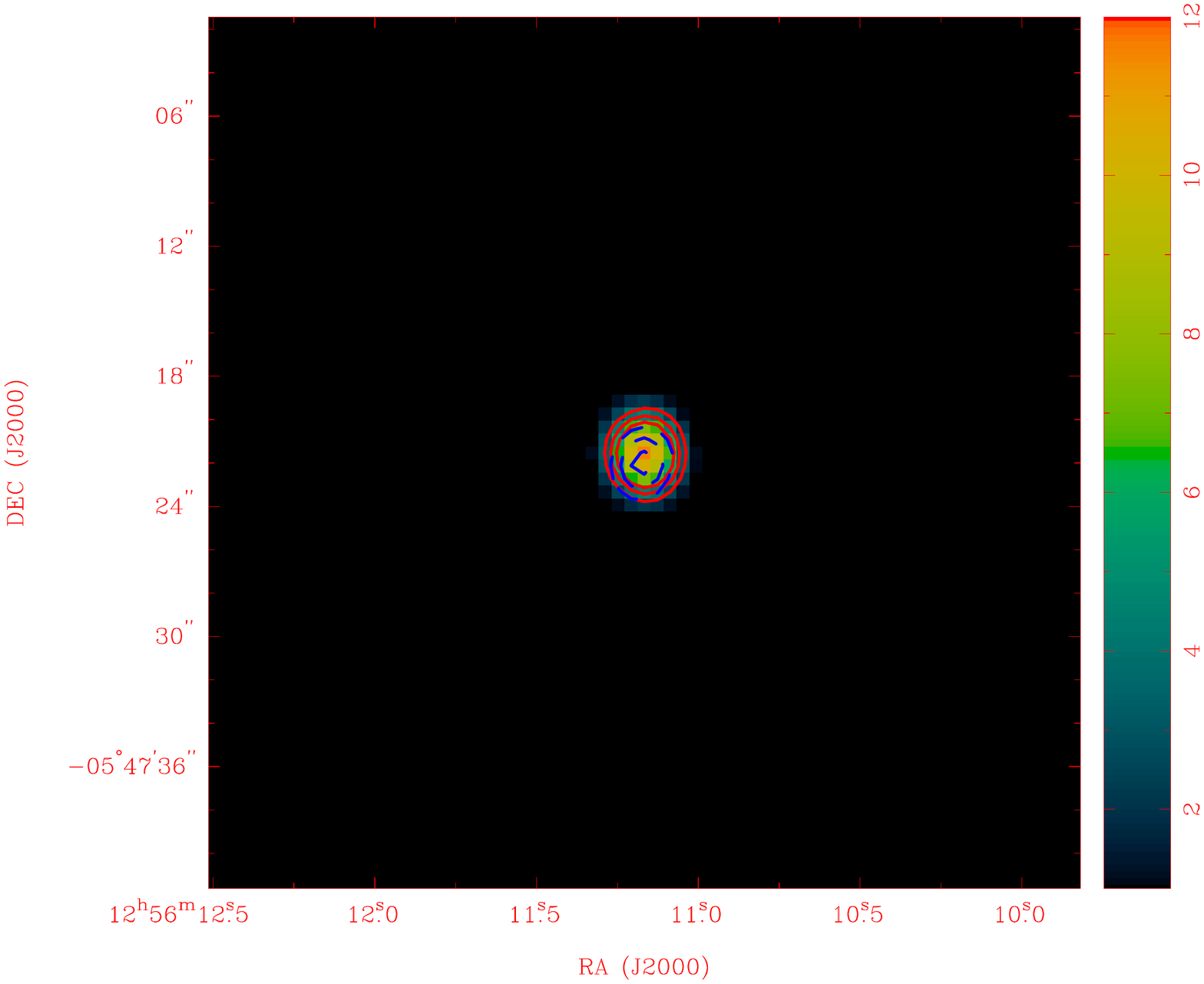}
\includegraphics[width=4.5cm,angle=-90]{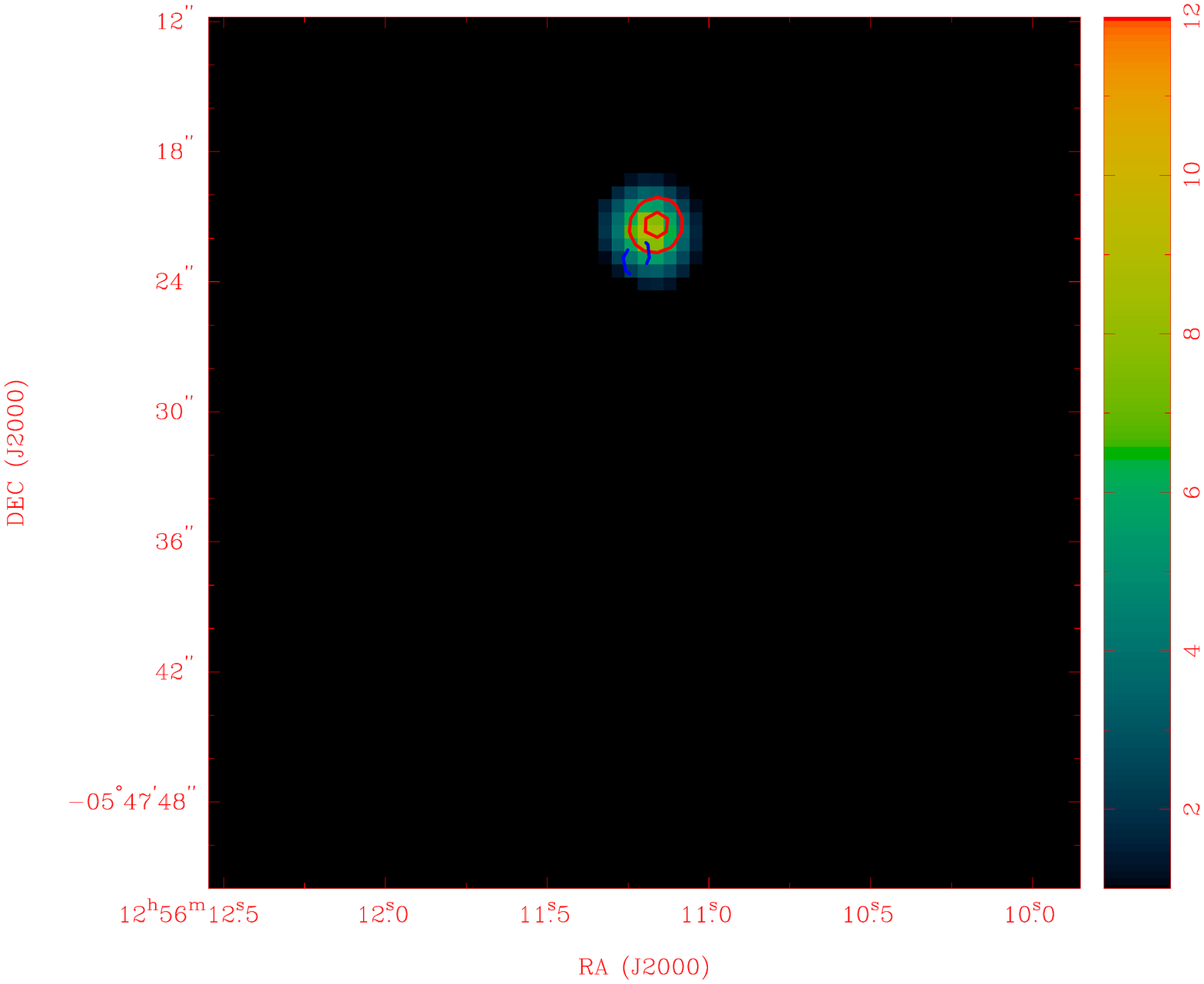}
\includegraphics[width=4.5cm,angle=-90]{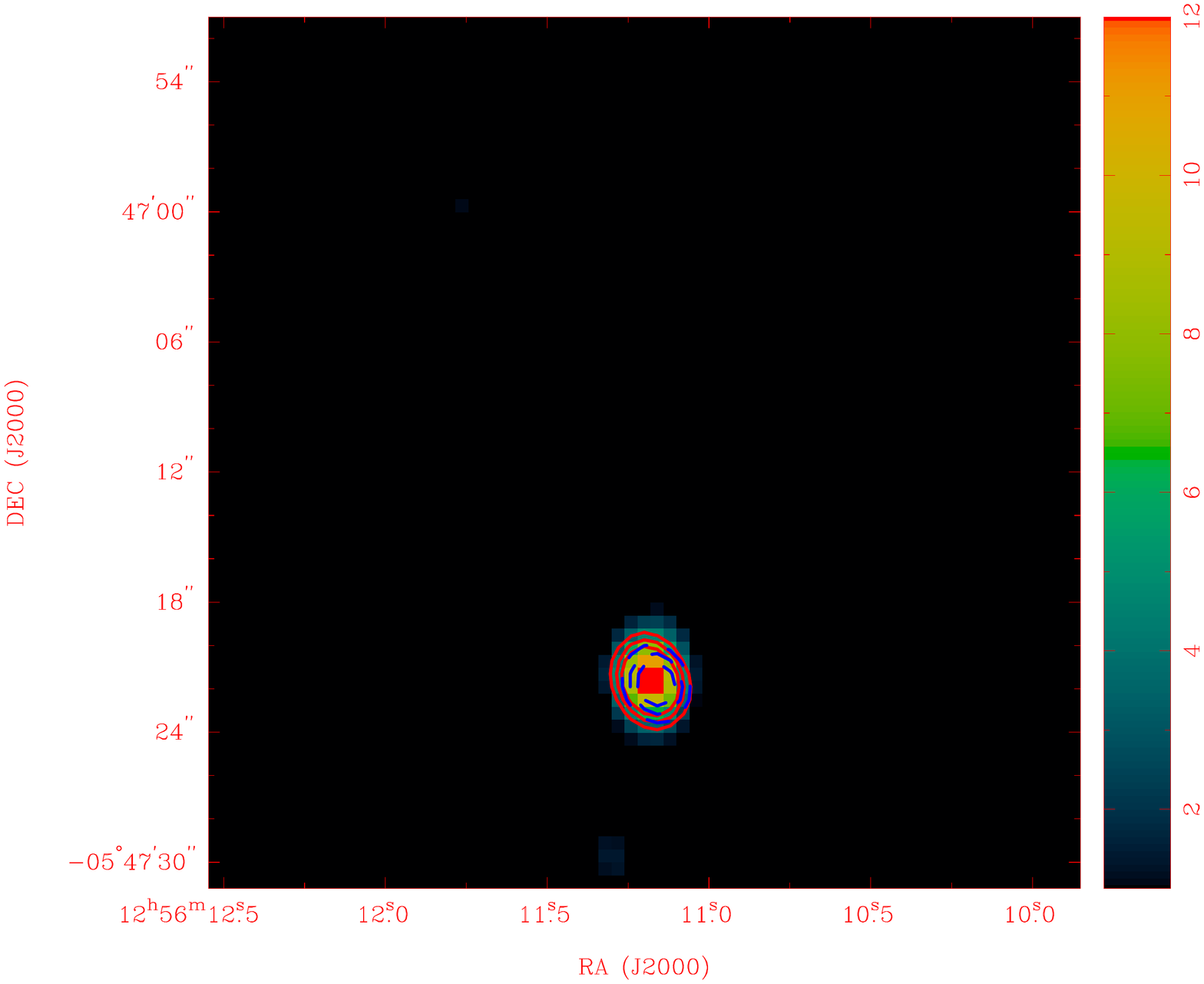}
\caption{\textit{Top}: Cleaned intensity map for the calibrator 3C279 and two of its offsets. The fractional polarisation amplitude image and the position angle image are together used to display the polarisation vector map. Using Miriad conventions, the vectors are scaled so that the maximum amplitude ($A_\mathrm{max}=0.03$) takes 1/20 of the plot size. The scale bar (below the beam FWHM at the bottom left corner) corresponds to the length of the longest vector. \textit{Bottom}: polarisation contours Q (red) and U (blue) overlayed on  intensity map for the calibrator 3C279 and two of its offsets.  The Q and U polarisation contour levels are at $-2,-1.5,-1,1,1.5,2$ Jy/beam for the calibrator and the two offsets. \label{offset}}
\end{figure*}

During the course of the first observation, we used twelve offset positions for the calibrator 3C279 with the aim to measure and correct for variations in instrumental polarisation across the telescope primary beams. 3C279, a bright radio quasar, is inherently polarised (typically at the 5-10\% level) and traces of anomalous polarisation signal if observed, could be only attributed to instrumental leakages. Unfortunately, due to an error on our part, the E/W coordinates of most of the offset pointings were misinterpreted during the execution of the observation script; only the observations of North-South axial offsets (offsets 8 and 11 as seen in Figure \ref{offset_pos}) turned out to be useful in their entirety. We did a naive mapping of the polarisation pattern for all these offsets and used it as a weighted template for correcting the instrumental polarisation leakages. In Figure \ref{offset} we show the maps for the calibrator and two of the offsets. 

To correct for the instrumental polarisation, we use the information from the twelve offsets of calibrator 3C279 as shown in Figure \ref{offset_pos}. Using the Miriad \textit{maths} task, we obtain weighted linearly combined Q and U maps for the 3C279 offsets field and use this as a model of the instrumental polarisation leakages. We use the combined offset field polarisation map and weight it with a factor proportional to the ratio of the \textit{rms} of the source and the calibrator as a polarisation leakage correction template. 

\subsection{Deconvolution issues}
In addition to instrumental polarisation leakages, other systematic effects could be of concern in the experimental setup. An important issue is deconvolution technique used in the CARMA software. We use a nine-point mosaic for Saturn, six along a inner hexagon, one at the center and 2 along the rings as shown in Figure \ref{mosaic}. However, standard data reduction demands that we use only the central pointing of Saturn during \textit{selfcal} and \textit{invert}. Thus the complete n-point mosaic  information is lost during the final stages. A reliable solution would be to extract the information from the various pointings.

We extract this information using the Miriad task \textit{demos}, which performs an inverse mosaicing operation for various pointings of Saturn for the various antenna configurations in the the E-array (OVRO, BIMA and their cross-correlations). We extract 27 pointings using \textit{demos} (9 mosaics of Saturn for each set of three antenna configurations) and perform \textit{selfcal} for each pointing. However the gain solutions do not match for the three sets of antennae configuration as we cannot have a common reference antenna for the three different antenna configurations. Due to presence of a phase structure in the primary beam outside the half-power radius, the gain solutions for different pointings do not match. In principle it is possible to only choose the OVRO (or the BIMA) antennae, for which the gain solutions would match for each pointing; however that would reduce the signal-to-noise ratio by a considerable factor and is hence not useful.

\begin{figure*}
\centering
\includegraphics[width=4.5cm,angle=-90]{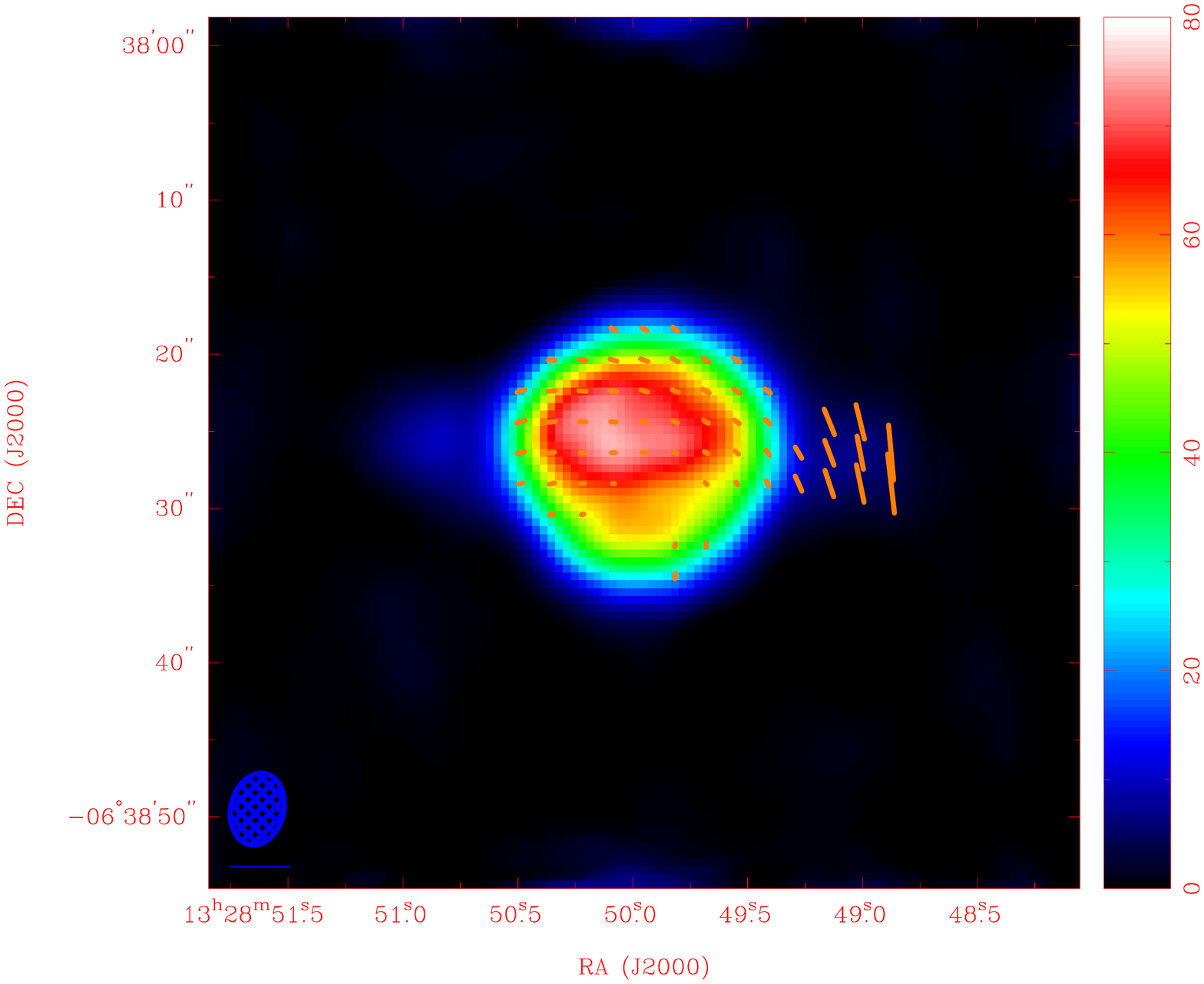}
\includegraphics[width=4.5cm,angle=-90]{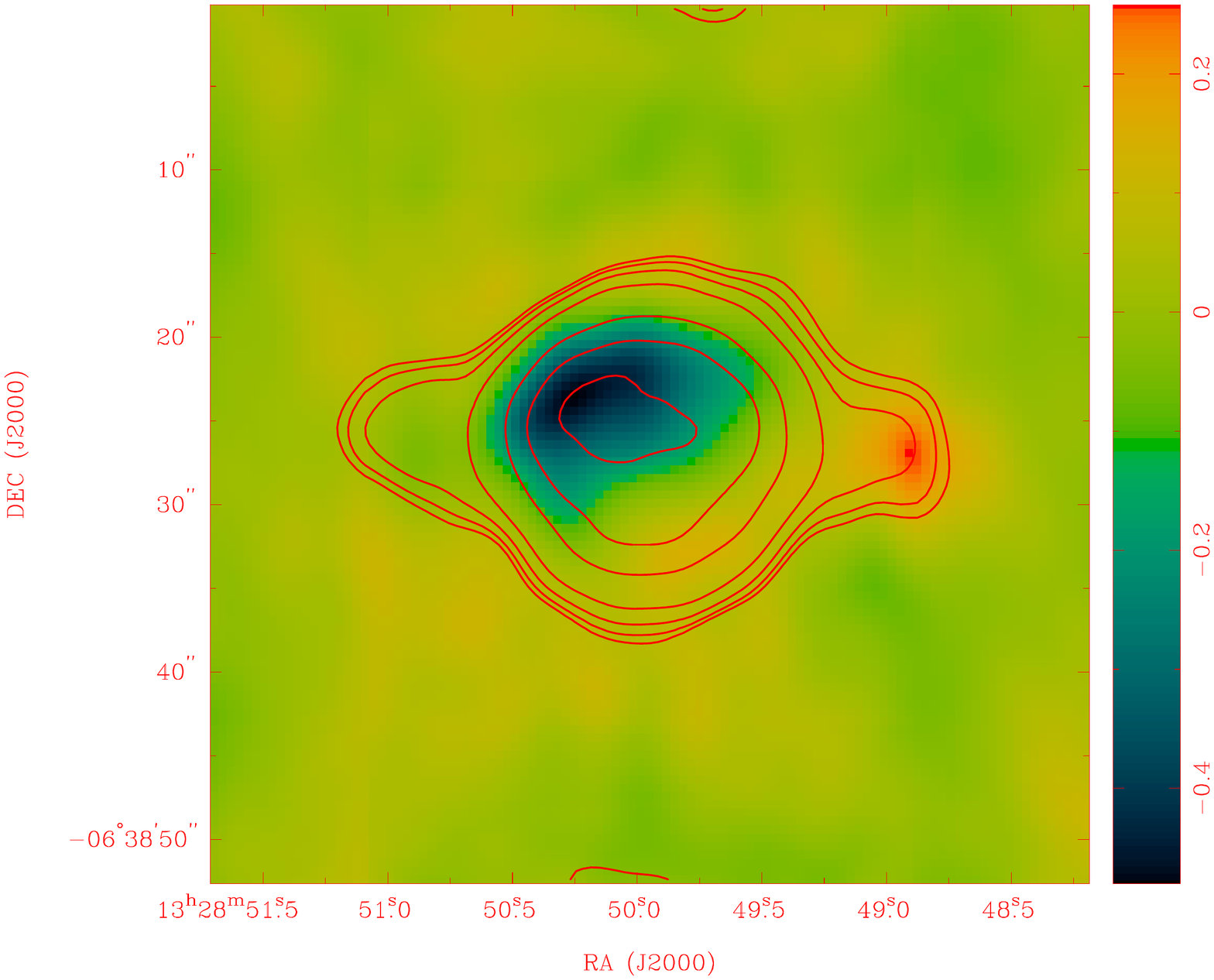}
\includegraphics[width=4.5cm,angle=-90]{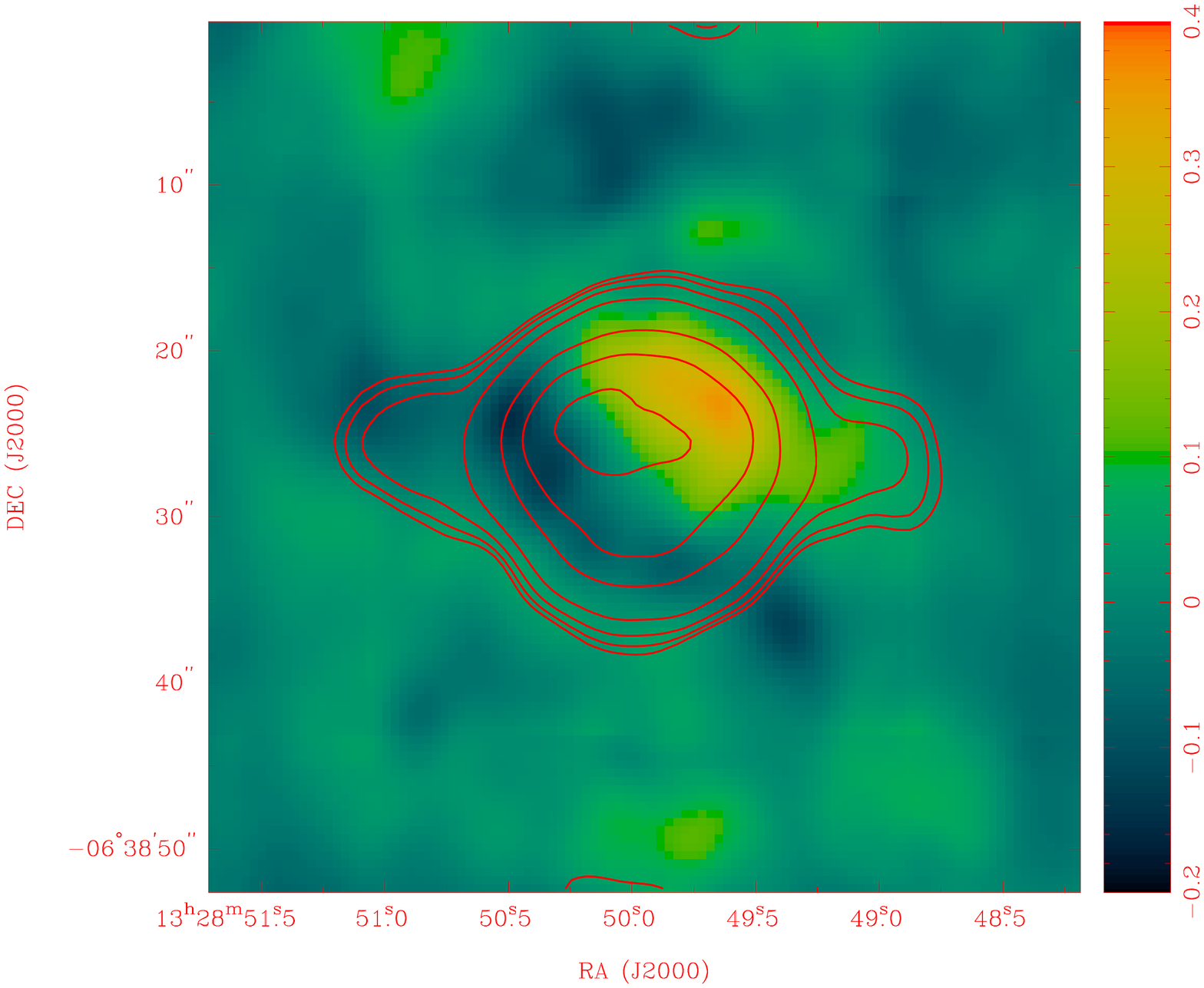}
\includegraphics[width=4.5cm,angle=-90]{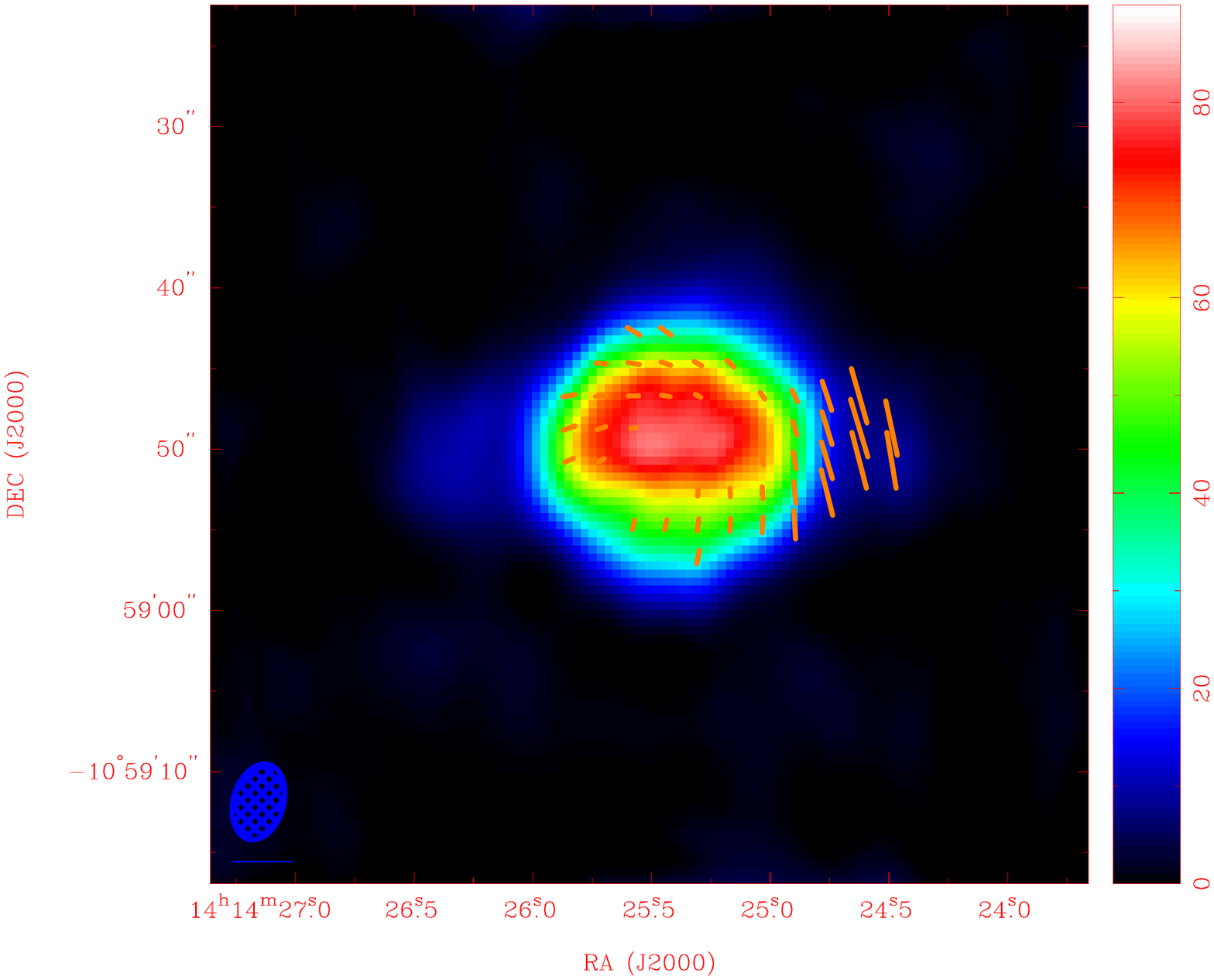}
\includegraphics[width=4.5cm,angle=-90]{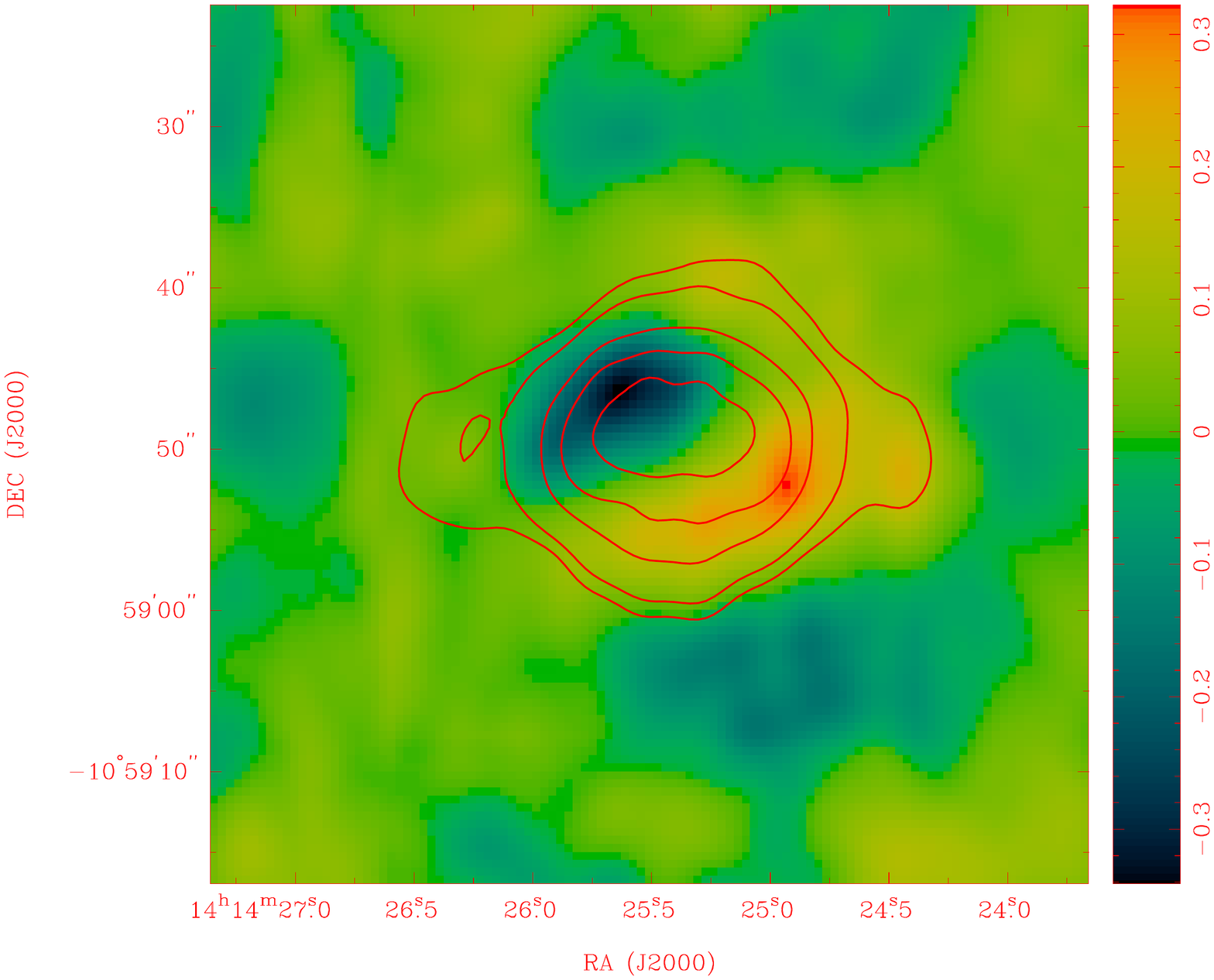}
\includegraphics[width=4.5cm,angle=-90]{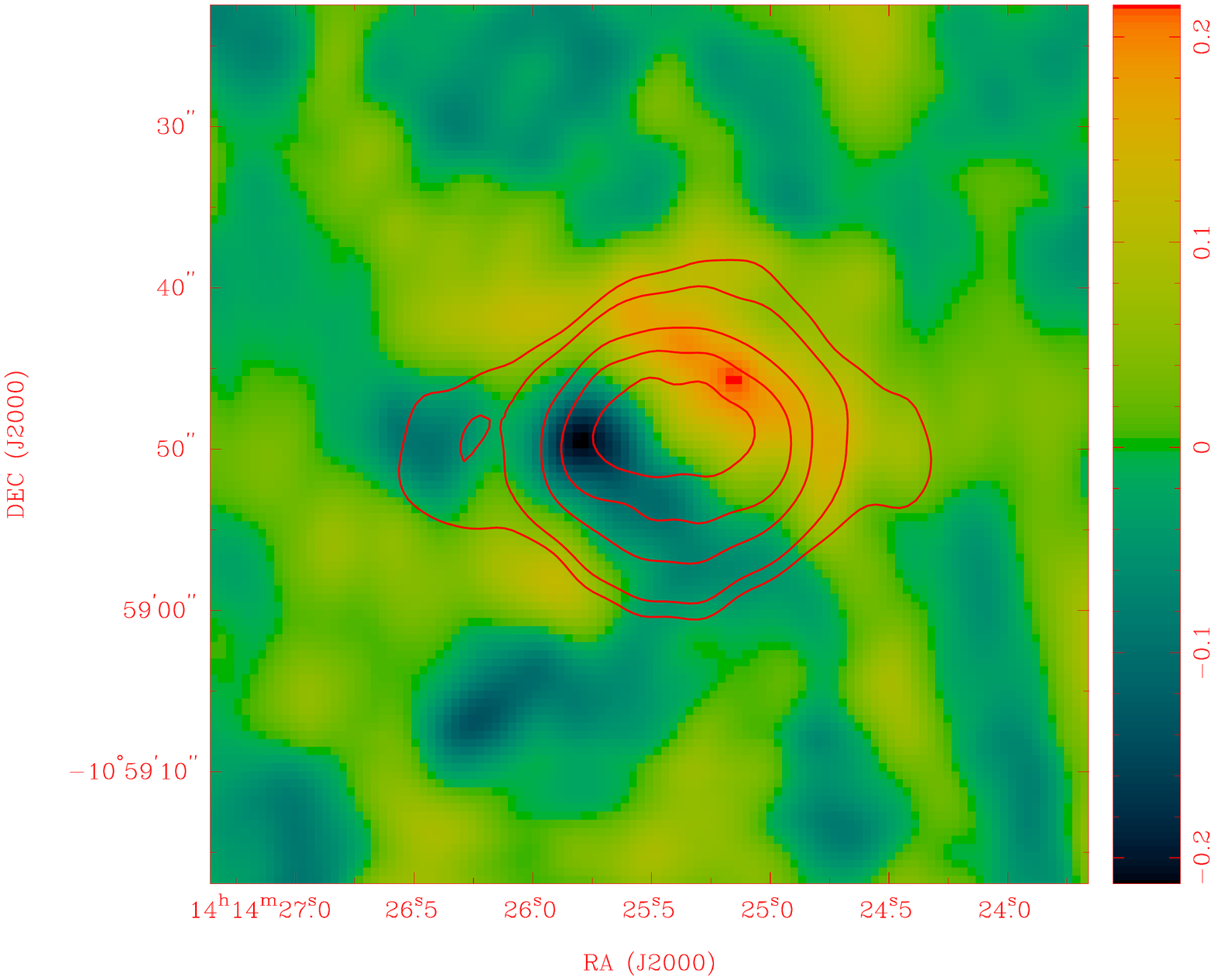}
\caption{\textit{Top}: Using dataset I, the intensity map of Saturn overlayed with polarisation vectors is shown (\textit{left}). The fractional polarisation amplitude image and the position angle image are together used to display the polarisation vector map. Using Miriad conventions, the vectors are scaled so that the maximum amplitude ($A_\mathrm{max}=0.06$) takes 1/20 of the plot size. The scale bar (below the beam FWHM at the bottom left corner) corresponds to the length of the longest vector. The Q (\textit{center}) and U (\textit{right}) pixel maps with intensity contours (contour levels are at $2,3,5,10,30,50,70,90$ Jy/beam) marking the position of Saturn's disk are also shown. These three figures are after correcting for instrumental polarisation and has negligible difference with the maps obtained without weighted noise correction. The asymmetric polarisation pattern is clear from these maps. \textit{Bottom}: Similar maps of Saturn using dataset II are shown, where the total intensity map of Saturn overlayed with polarisation vectors has maximum amplitude ($A_\mathrm{max}=0.03$). The intensity contours on the Q and U pixel maps are at same values as dataset I maps. The asymmetric polarisation pattern is also clear from these maps. }
\label{sat_pol}
\end{figure*}

\begin{figure*}
\centering
\includegraphics[width=6.8cm,angle=-90]{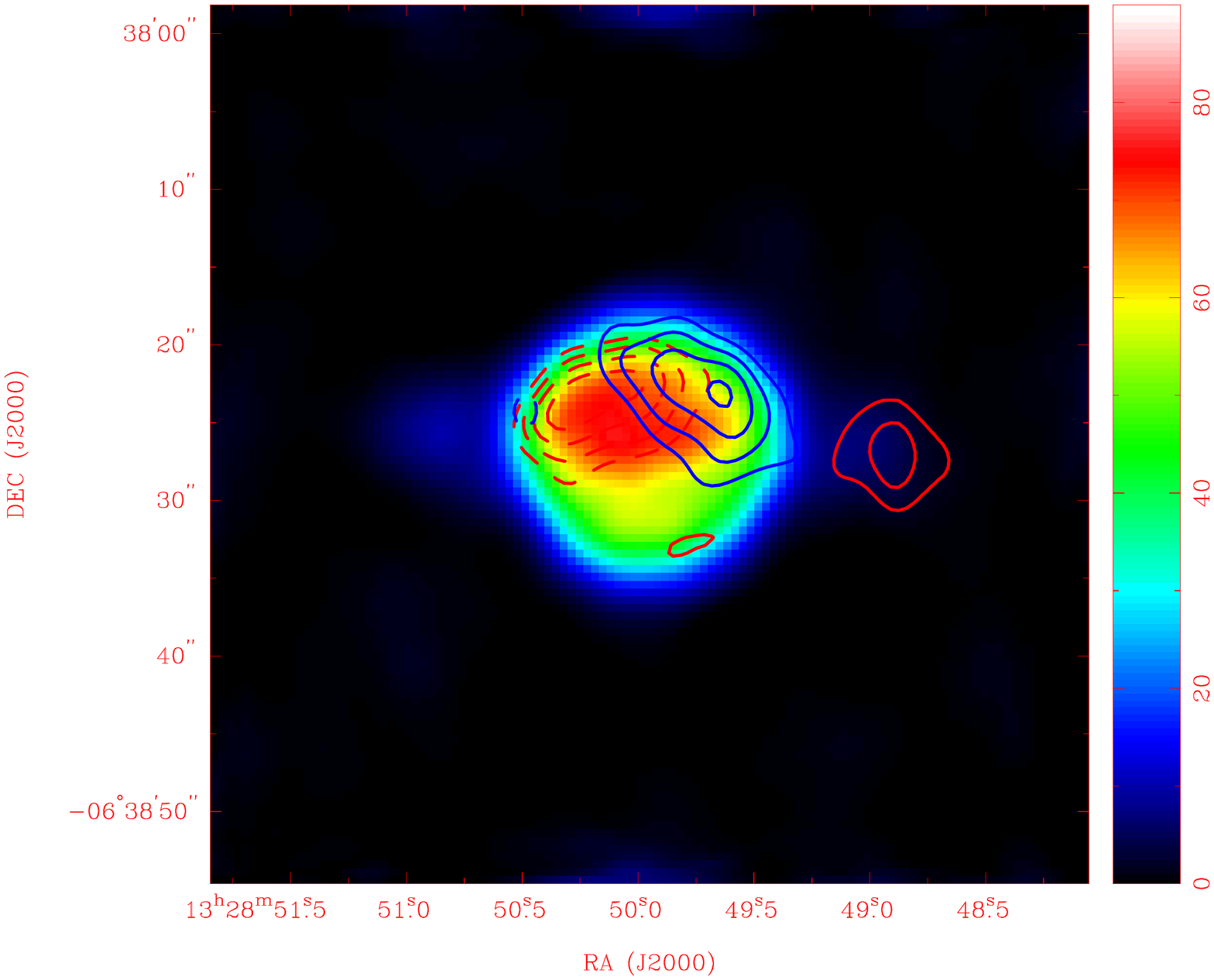}
\includegraphics[width=6.8cm,angle=-90]{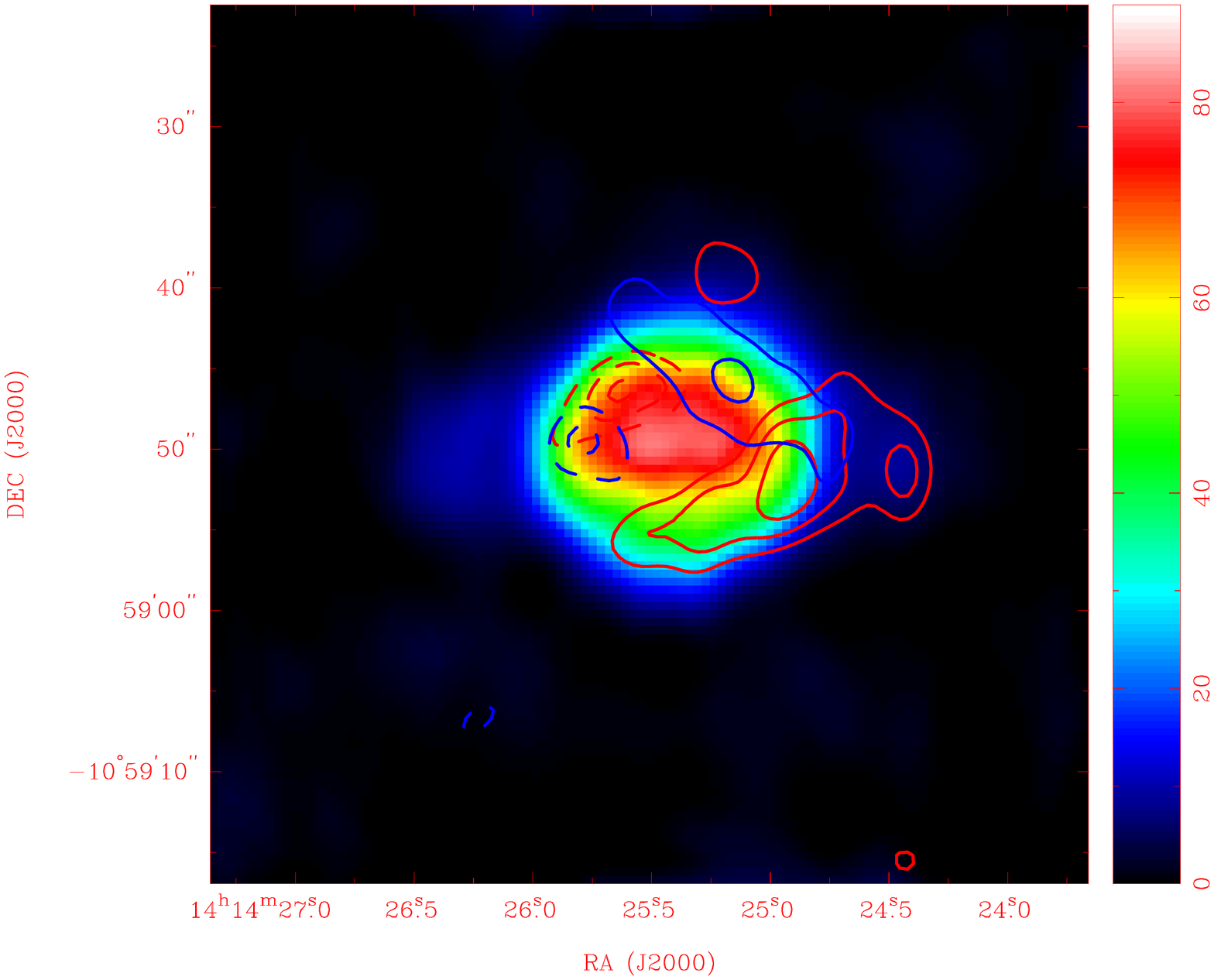}
\caption{ \textit{Left}: Observation set I - Q (red) and U (blue) polarisation contours overlayed on intensity map of Saturn. The contour levels are at $\{-0.42,-0.35,-0.28,-0.21,-0.14,0.14,0.21,0.28,0.35,0.42\}$ Jy/beam for both Q and U contours. 
\textit{Right}: Observation set II - Q (red) and U (blue) polarisation contours overlayed on intensity map of Saturn for dataset II with same contour levels as dataset I. }
\label{saturn2}
\end{figure*}

\begin{figure*}
\centering
\includegraphics[width=6.8cm,angle=-90]{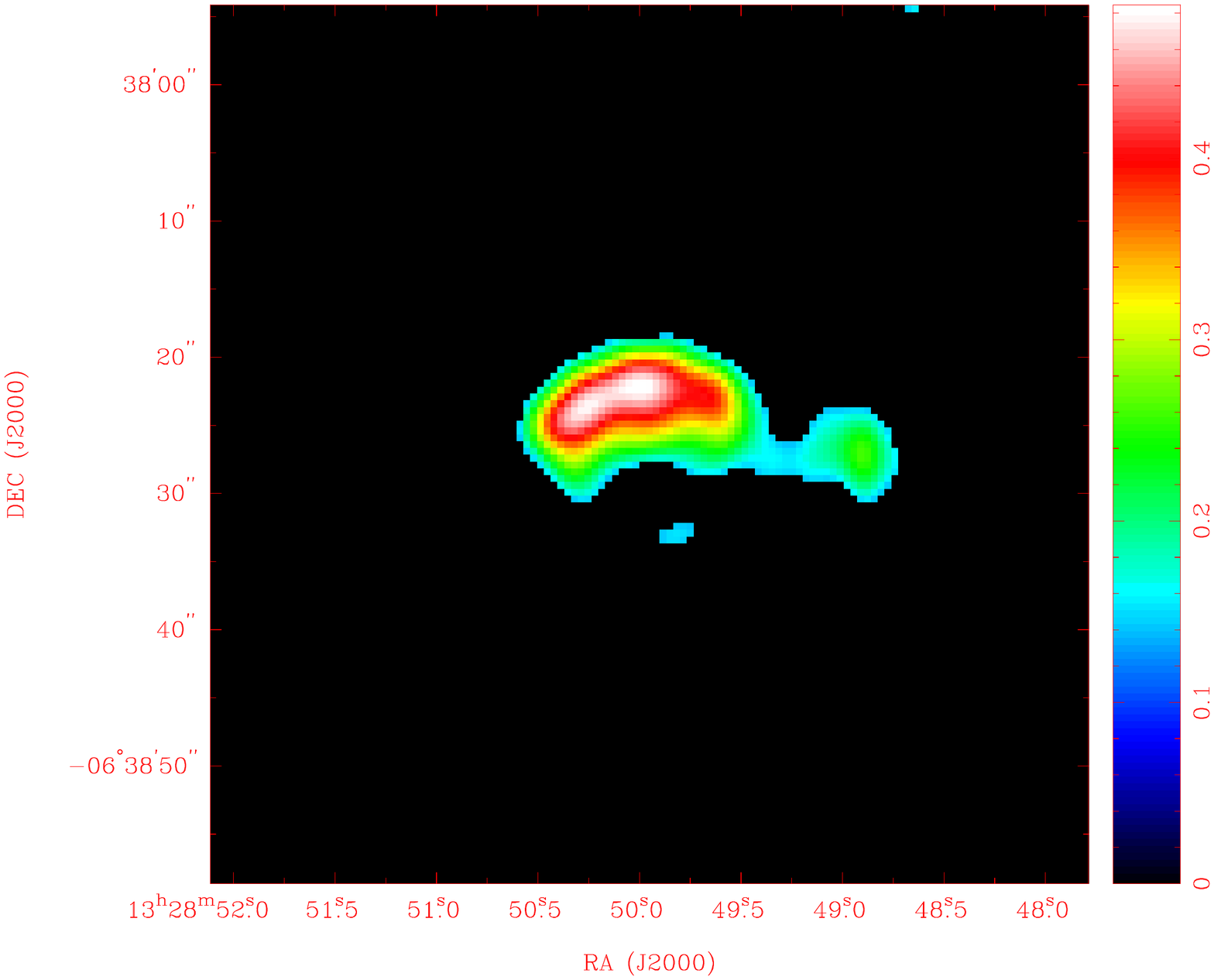}
\includegraphics[width=6.8cm,angle=-90]{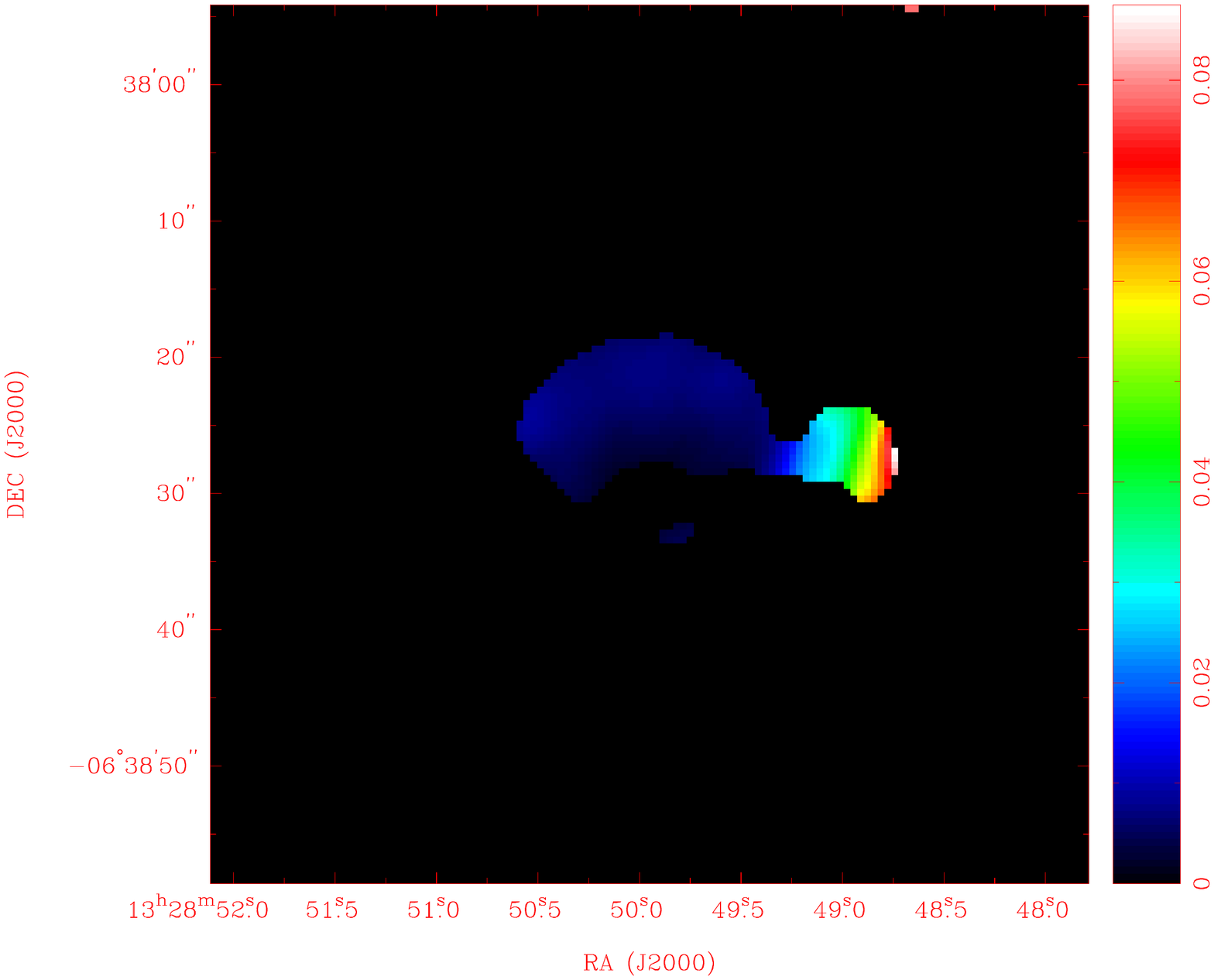}
\caption{Observation set 1 - \textit{Left}: Total polarisation intensity map of Saturn, with a range of $0.1 - 0.5$ Jy/beam. \textit{Right}: Fractional polarisation map of Saturn showing $\sim 9\%$ fractional polarisation on the west side of the rings. The corresponding pixel maps for dataset II for Saturn is similar.}
\label{poli1}
\end{figure*}

\begin{figure*}
\centering
\includegraphics[width=6.8cm,angle=-90]{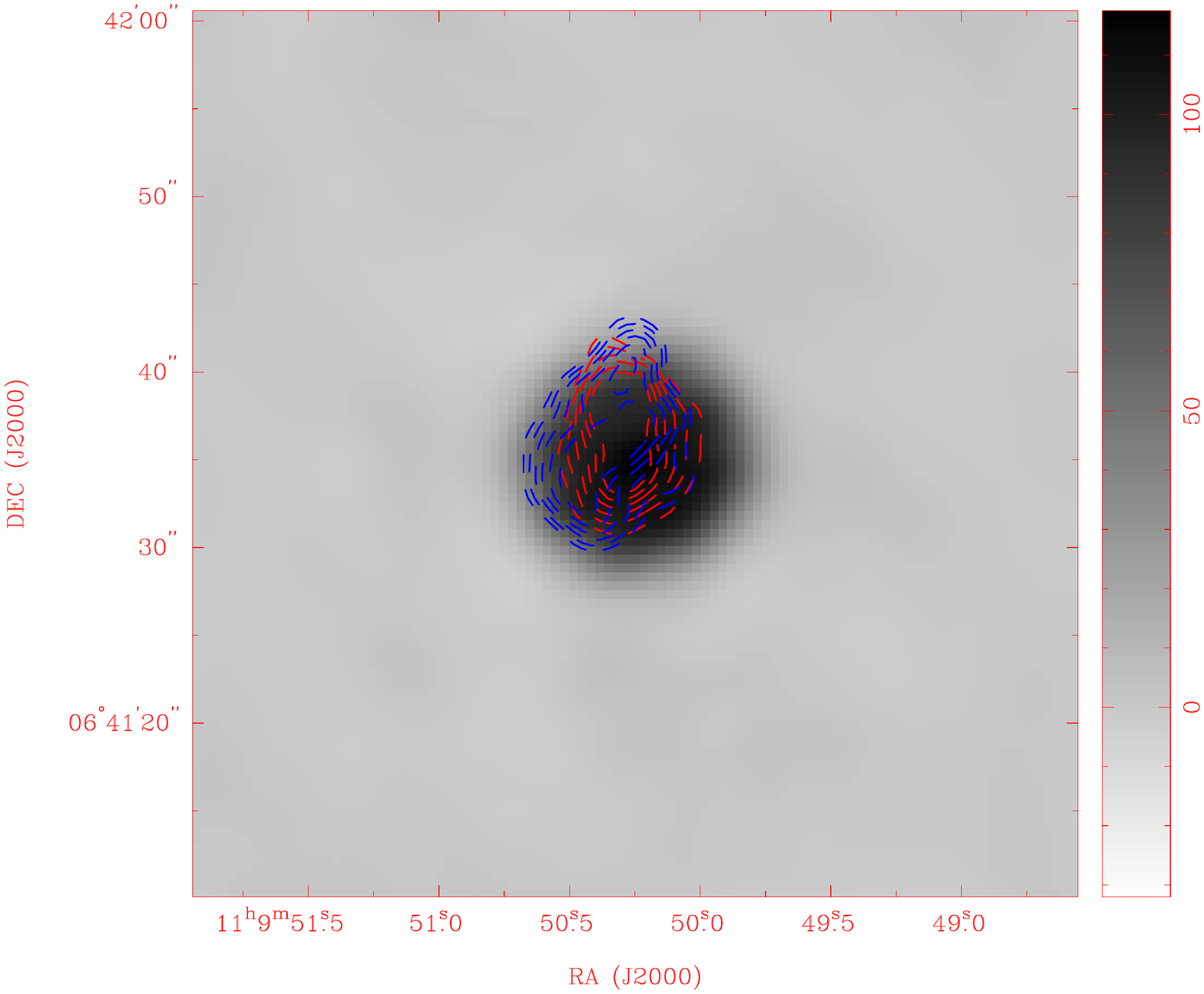}
\includegraphics[width=6.8cm,angle=-90]{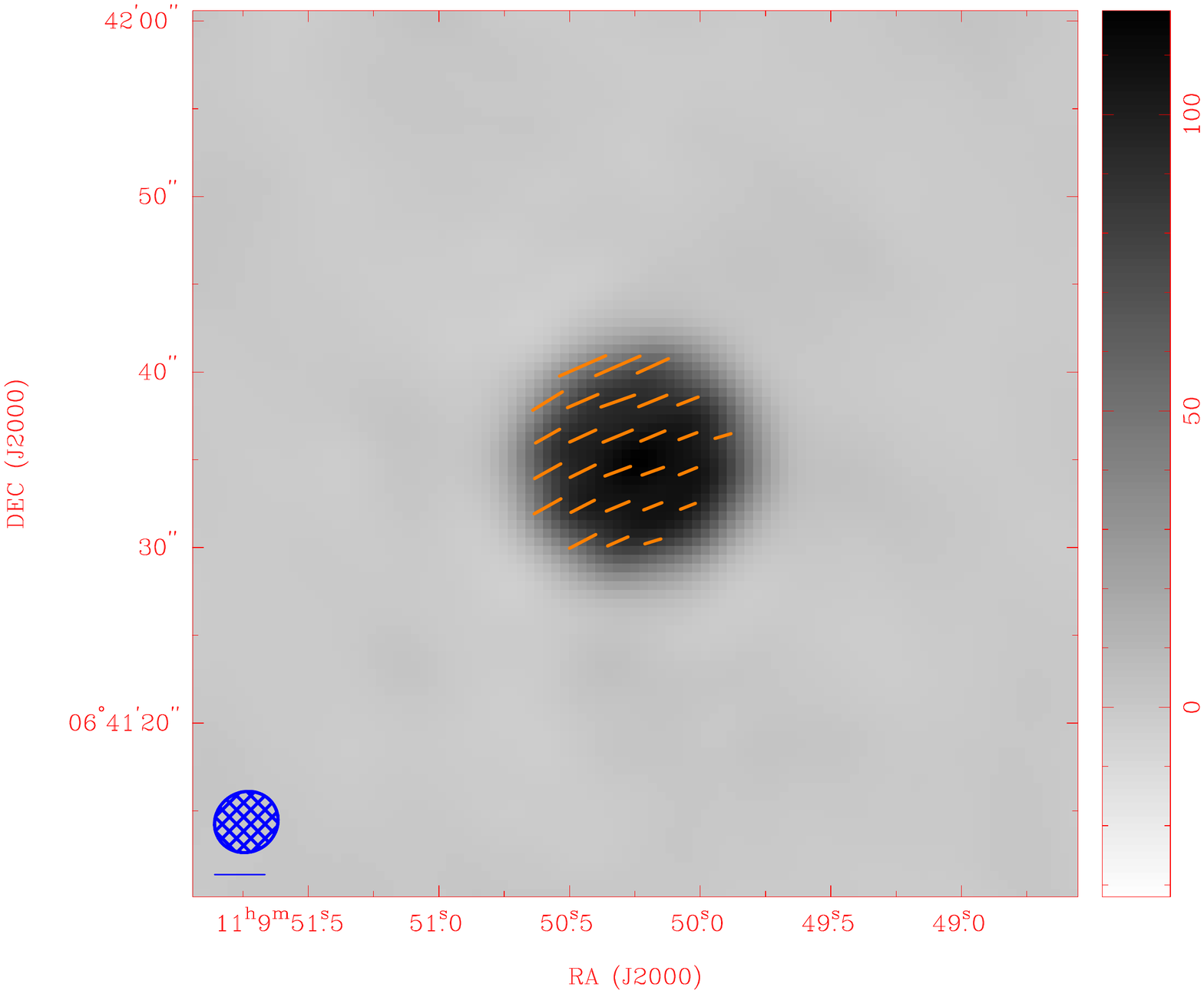}
\caption{ \textit{Left}: Q (red) and U (blue) polarisation contours overlayed on intensity map of Venus. The contour levels are at $\{-1.5,-1.4,-1.3,\, ...\, ,1.3,1.4,1.5\}$ Jy/beam for both Q and U contours. \textit{Right}: Intensity map of Venus overlayed with polarisation vectors. The fractional polarisation amplitude image and the position angle image are together used to display the polarisation vector map. Using Miriad conventions, the vectors are scaled so that the maximum amplitude ($A_\mathrm{max}=0.047$) takes 1/20 of the plot size.} 
\label{venus}
\end{figure*}

\section{Results}
We discuss the results obtained using the first dataset as in section \ref{obs1}. The top panel in Figure \ref{sat_pol} shows Saturn's intensity map overlayed with the polarisation vectors and the Q/U pixel maps after the leakage correction as discussed in subsection \ref{leakage-corr}. It is interesting to note that there is 1.5\% fractional polarisation on the disk of Saturn, matching with results from \cite{grossman}, \cite{dunn}. However, the puzzling feature we saw was the asymmetric 9\% fractional polarisation only on the west side of the rings of Saturn. The effect of instrumental leakage correction using the calibrators is negligible. The leakage template obtained thus is not sufficient to explain the anomalous polarisation distribution along the rings. 

We aimed to resolve this anomalous polarisation pattern with the second dataset, described in section \ref{obs2}. Observed closely one year apart, we however find the final maps of Saturn show the exact same asymmetric polarisation pattern on the west side of the rings.  The bottom panel of Figure \ref{sat_pol} shows similar I, Q, U maps for dataset II as discussed in section \ref{obs2} which records slightly lower temperature and polarisation intensity. However, the asymmetry on the west side of the rings is clear in the polarisation vector maps. Due to lack of calibrator offset observations, we fail to perform a similar leakage correction using the weighted offset maps as for dataset I. 

Figure \ref{saturn2} shows the polarisation contours for Q (in red) and U (in blue) on the intensity map for dataset I (left) and dataset II (right). The solid contours are for positive Q/U while the dashed contours represent negative Q/U. This also clearly shows that there is high polarisation asymmetric on the east and west side of the rings. 

The pixel maps in Figure \ref{poli1} shows the total polarisation intensity and the fractional polarisation respectively for dataset I of Saturn. It is seen from the first dataset, the fractional polarisation on the west side of the ring is close to 9\% as seen on the right of Figure \ref{poli1}. The corresponding pixel maps for dataset II for Saturn is similar.

In Figure \ref{venus}, we see similar polarisation patterns on Venus, which in this case could be interpreted as partly instrumental. However the maps on Figure \ref{venus} lack the asymmetric pattern as seen along the rings of Saturn. Limitations of CARMA from dynamic range of polarisation measurement for bright sources which are weakly polarised could explain spurious polarisation pattern in the images; however the `east-west' asymmetry in the polarisation pattern on Saturn's rings remains unexplained in this context. 

At 1-3 mm, the primary source of flux from the rings should be the ring's own thermal emission, and reflected Saturn signal should be about 10\% of the total brightness (See Figure 7 of \cite{dunn}). At 2-6 cm-wavelengths (where the primary signal is reflected Saturn-signal) and at high opening angles, the polarisation of the ring is around 20\% on both ansa and roughly North-south (see \cite{grossman} and \cite{tak}). This suggests that pure scattered Saturn signal would create vertically polarised signals like the one we see in Figure \ref{sat_pol}. In terms of our analysis, what needs explaining is why the signal is so asymmetric between the two ansa and why the signal on the right ansa is so strong.

\subsubsection*{Instrumental polarisation}
We were concerned whether instrumental polarisation leakage was masquerading as this asymmetric polarisation signal. In order to eliminate the instrumental polarisation leakages, we use the combined information from the calibrator 3C279 and its offsets as discussed in subsection \ref{leakage-corr}. Although instrumental polarisation leakage could in principle give rise to small fraction of asymmetry in the polarisation pattern, it does not explain the large asymmetric 9\% fractional polarisation signal on the west side of the ring as seen in Figure \ref{sat_pol}, \ref{saturn2} and \ref{poli1}. 

To check the levels of instrumental polarisation leakages in CARMA E-array, we used standard polarisation scripts in Miriad (\$MIR/demo/carma/stokes.csh) to simulate CARMA's polarisation leakages. These scripts are tailored for the CARMA array and we feed in polarisation leakages by hand and study the distortion in the u-v space and also in the map space. We simulate the polarisation contours due to a point source at the center without any polarisation leakage. We compare this with another simulation where we have a Gaussian beam at the center. The FWHM of the beam is manually adjusted to match Saturn's polarisation intensity obtained during the data analysis. We check the effect of fractional instrumental polarisation leakage due to this Gaussian beam at the center and find that a high polarisation leakage fraction of more than 10\% is required to give rise to anomalous Q/U contour patterns. We can thus conclude from the simulation that it would require an unnaturally large polarisation leakage fraction to reproduce the polarisation asymmetry as shown in Figures \ref{sat_pol}, \ref{saturn2} and \ref{poli1}. 

We briefly discuss the accuracy constraints on polarisation measurements in subsection \ref{dynamic}. Since CARMA uses circularly polarised feeds for dual polarisation, leakage ripples are higher than linear polarised feeds, which leads to introducing frequency dependent cross-coupling between the two polarisation channels. Additionally, beam squint could also cause an offset in the LCP and RCP responses. Squint exists in reflector with circular symmetric when the circularly polarised feed is laterally offset from the focus and, more significantly, in offset systems regardless of whether the circularly polarised feed is on or off axis. For CARMA, the feeds on the 10m array are on-axis but odd number of off-axis multiple reflections from mirrors preceding the receiver, causes a beam squint which is worse compared to the 6m array where the feeds are a little offset. These are issues which has been studied in details recently in \cite{hull}.

While the polarisation on Saturn's disk (Figure \ref{sat_pol}, \ref{saturn2}, \ref{poli1}) and on Venus \ref{venus} could have a contribution from polarisation leakages in the instrument, the asymmetry in the polarisation pattern on the rings of Saturn (Figure \ref{sat_pol}, \ref{saturn2}, \ref{poli1}) is relatively difficult to explain through instrumental leakages alone. This strong `east-west' polarisation asymmetry on the rings could have a component arising due to instrumental leakages; however to explain the 9\% fractional polarisation on the west side of the rings, compared to the negligible fractional polarisation on the east side (which has relatively equal intensity compared to the west side) needs further analysis with future high quality polarisation data from the Atacama Large Millimeter/submillimeter Array (ALMA). 

\subsubsection*{Possibility of self-gravity wakes}
A possible explanation could be that this asymmetry actually rises from some physical phenomenon. It is well known that in parts of the A and B rings, the particles are organized into elongated aggregates known as self-gravity wakes \citep{colwell}. Self-gravitating wakes could tend to give a polarisation. The wakes tend to be in the A ring and trail rotation by about 22$^\circ$ as seen in Figure 5 in \cite{dunn-2004}. Light, coming from Saturn would be more likely to bounce of the \textit{west} side wakes and perhaps pick up more polarisation than the \textit{east} side wakes, since light would have to transmit through the wakes more. 

For a simply (non wakey) ring, we would expect the thermal emission from the ring to not be strongly polarised at high opening angles. This is because the thermal emission from a surface is polarised in the direction of the surface normal, and for quasi-spherical particles should average to zero. However, for the CARMA observations, the elevation angle of the rings is only about 13 degrees. In that case ring particles can block one another, and we might get something more like a sheet that is tilted away from us. In that limit, the polarisation should again be N-S, as we see.

Next we try to explain the asymmetry. It should be noted that the opening angle we are using are where the asymmetries in the filling factor are expected to be the strongest (see Figure 3 of \cite{salo} and \cite{french}). Previous work which indicated an asymmetry attributed to gravity wakes was seen with radar observations made at Arecibo Observatory by \cite{nicholson2005}. Given that thermal emission from a spherical object forms a radial pattern, we would expect that the polarisation from thermal emission of wakes would be perpendicular to the wakes. However, by Kirchhoff laws, the reflected Saturn-shine would be polarised parallel with the wakes. 

A key point for why self-gravity wakes could give a polarisation asymmetry is that the rings should be relatively strongly back-scattering. Parts of the rings behind the planet i.e. from the distant half of each ring ansa, should preferentially contribute to the scattered Saturn-shine signal. 
This would produce an E-W signal on one ansa and a N-S on the other which would be seen as a strong asymmetry in the total polarisation fraction. The scattered signal on the east ansa is primarily horizontally polarised, but is more vertically polarised on the west ansa (see Figure 12 of \cite{nicholson2005}).

We suspect that we are seeing a combination of three phenomena: N-S polarisation from the regions where the wakes are seen ``side on'' and form an opaque ring, E-W polarisation from the wake thermal emission where we wakes are being seen ``end on'', and N-S polarisation from the side-scattered Saturn light. On the right ansa we can see the scattered Saturn-shine because the inner side of the wake is tilted towards us, while on the other ansa that signal is largely blocked from us.

We are seeing mostly ring thermal emission on the east ansa, and that the various polarised signals from the wakes are canceling each other out, but on the west ansa the extra signal from the planet's thermal emission is allowing there to be a net polarisation. However, to actually see if this model works quantitatively, we would need to do a proper radiative transfer calculation, probably using Dunn's simrings code \citep{dunn}, which is something we are planning to work on. 

\section{Conclusion} 
We have observed Saturn in polarisation twice separated by a period of 10 months. Apart from the the difference in fractional polarisation, there is no change in the pattern of the polarisation in both the observations. We have checked that the asymmetric polarisation pattern we see on Saturn's rings, could be a contribution from the instrumental polarisation leakage. However, the polarisation signal on the west side of the ring is higher than expected from only instrumental polarisation leakage. We have tried to correct for the possible polarisation leakage, using a template developed using twelve calibrator offsets where we perform a weighted average of the polarisation leakage field using a linearly combined weighted polarisation map of the calibrator offsets.  We take into account this correction which however is not sufficient to explain the anomalous polarisation signal seen only on the west side of the planet's rings. 

We find similar polarisation pattern on Venus which could have spurious contributions due to the stretching in the realms of the dynamic range of the telescope. However the asymmetry in the polarisation on Saturn's rings are not fully explained by the dynamic range limitation. But as discussed above, these effects are a serious concern for weakly polarised sources; in our case the maximum of polarisation to intensity ratio is about a percent, which makes a strong case for dynamic range limitation. The reason for these inconsistencies from dataset to dataset is not fully understood. Given the instrumental constraints for polarisation measurements in CARMA, we cannot concretely confirm the effect of self-gravity wakes on the asymmetric polarisation pattern seen on Saturn's ring system. This study thus stands as an upper constraint of polarisation limits on Saturn's disk and ring system at 220 GHz using CARMA E-array configuration. We plan to extend our observations and modeling of the polarisation pattern of Saturn using ALMA when polarisation for extended sources will be commissioned, hopefully early 2016 for ALMA Cycle 4 observations. 

\section*{Acknowledgment}
MA would like to thank the instructors of the CARMA summer school 2013, especially Richard Plambeck, Melvyn Wright and Nikolaus Volgenau. The school was a great opportunity to learn the intricacies of radio data analysis from experts in the field. All the comments and directions were very constructive in understanding our observations. We also thank Danielle Lucero for useful discussions. 

\bibliographystyle{mnras}
\bibliography{saturn}

\label{lastpage}

\end{document}